\documentclass[a4paper]{article} 
\title{Predicting the Curie Temperature of Magnetic Materials with Machine Learning: Descriptor Engineering, Graph Neural Networks, and the Role of Curated Data}
\usepackage{makecell}

\usepackage{dblfloatfix}
\usepackage{changepage}
\usepackage{indentfirst}
\usepackage{ragged2e}
\usepackage{longtable}
\usepackage{adjustbox}
\usepackage{etoolbox}
\usepackage{multicol}
\usepackage[justification=raggedright,singlelinecheck=false]{caption}
\usepackage{authblk}

\usepackage[english]{babel}
\usepackage{array} 
\usepackage{float}  
\usepackage{amsmath}
\usepackage{graphicx}
\usepackage[left=1.2cm, right=1.2cm, top=1.5cm, bottom=1.6cm]{geometry}
\usepackage[utf8]{inputenc}
\usepackage[colorlinks=true, linkcolor=blue, citecolor=blue, urlcolor=blue]{hyperref}
\usepackage[backend=biber,style=numeric,sorting=none]{biblatex}
\usepackage{authblk}
\usepackage{xcolor}

\usepackage[normalem]{ulem}
\usepackage{etoolbox} 

\makeatletter
\newcommand{\redsout}[1]{
  \begingroup
    \color{red}
    \def\UL@color{\relax}
    \sout{#1}
  \endgroup
}
\makeatother
\author[1]{Akram Abedi Orang}
\author[1,2,*]{Mojtaba Alaei}
\author[2] {Artem R. Oganov}
\affil[1]{Department of Physics, Isfahan University of Technology, Isfahan 84156-83111, Iran}
\affil[2]{Skolkovo Institute of Science and Technology, Bolshoy Boulevard 30, bld. 1, Moscow 121205, Russia}
\affil[*]{Corresponding author: m.alaei@iut.ac.ir}
\begin{document}
\maketitle

\begin{center}
\noindent
\begin{minipage}[c]{0.93\textwidth} 
    \vspace{-25pt} 
    \hspace{0.9cm}
    \noindent

Predicting the Curie temperature ($T_\mathrm{C}$) of magnetic materials is crucial for advancing applications in data storage, spintronics, and sensors. We present a machine learning (ML) framework to predict $T_{\mathrm{C}}$ using a curated dataset of 2,500 ferromagnetic compounds, employing two types of elemental descriptor-based features: one based on stoichiometry-weighted descriptors, and the other leveraging Graph Neural Networks (GNNs). CatBoost trained on the stoichiometry-weighted descriptors achieved an $R^2$ score of 0.87, while the use of GNN-based representations led to a further improvement, with CatBoost reaching an $R^2$ of 0.91, highlighting the effectiveness of graph-based feature learning. We also demonstrated that using an uncurated dataset available online leads to poor predictions, resulting in a low $R^2$ score of 0.66 for the CatBoost model. We analyzed feature importance using tools such as Recursive Feature Elimination (RFE), which revealed that ionization energies are a key physicochemical factor influencing $T_\mathrm{C}$. Notably, the use of only the first 10 ionization energies as input features resulted in high predictive accuracy, with $R^2$ scores of up to 0.85 for statistical models and 0.89 for the GNN-based approach.
These results highlight that combining robust ML models with thoughtful feature engineering and high-quality data, can accelerate the discovery of magnetic materials. Our curated dataset is publicly available on GitHub.
\vspace{0.5cm} 

\end{minipage}
\end{center}
\begin{multicols}{2}
\section{Introduction}
Understanding and predicting the magnetic properties of materials, particularly the Curie temperature ($T_{\mathrm{C}}$),  
remains a fundamental challenge in materials science. The Curie temperature marks the transition of ferromagnetic materials to a paramagnetic state, controlled by complex quantum mechanical interactions, including spin exchange and magnetic anisotropy \cite{coey2010magnetism}. Ferromagnetic materials, play a central role in various technological applications. Accurate prediction of $T_{\mathrm{C}}$ can accelerate the design of magnetic materials, enhance the performance and reliability of magnetic devices, and reduce the experimental burden of material synthesis.

While first-principles approaches such as density functional theory (DFT) combined with effective Hamiltonians, typically the Heisenberg model, have been widely employed to estimate $T_{\mathrm{C}}$, they often suffer from high computational cost, limited scalability, dependence on structural data, and a strong dependence on the choice of exchange-correlation approximation
\cite{turek2006exchange, mosleh2023benchmarking, rezaei2025evaluating}.
In this context, machine learning (ML) has emerged as a transformative tool in materials science. By uncovering hidden patterns and correlations in large datasets, ML enables rapid and accurate predictions of materials properties, including thermodynamic stability, band gaps, elastic moduli, interatomic potentials, and magnetic transition temperatures \cite{schmidt2017predicting, zhuo2018predicting, nguyen2022machine, bartok2018machine, long2021accelerating}.

After the rise of ML, several independent studies have applied it to predict magnetic transition temperatures. Nelson and Sanvito developed a framework using only chemical composition, achieving a mean absolute error (MAE) of 57 K with a Random Forest model trained on ~2,500 compounds \cite{nelson2019predicting}.
Similarly, support vector regression (SVR) models showed strong performance in predicting Néel temperatures ($T_{\mathrm{N}}$) of antiferromagnets using atomic and chemical descriptors \cite{lu2021machine}.
A physics-informed ML model trained on Ce-based compounds achieved high accuracy ($R^2 \approx 0.95$, MAE $\approx$ 59 K) and was validated on Ce–Zr–Fe systems. DFT analysis further revealed correlations between $T_{\mathrm{C}}$, the density of states at the Fermi level, and the de Gennes factor in rare-earth intermetallics, underscoring the role of electronic structure in magnetic exchange interactions \cite{singh2023physics}.
In another comparative study, Random Forest outperformed k-nearest neighbors (k-NN) when trained on 2,500 compounds and tested on 3,000 additional entries, identifying cobalt-rich materials as having the highest $T_{\mathrm{C}}$ and iron-rich systems as cost-effective alternatives \cite{belot2023machine}.
Jung et al. applied a gradient-boosted pipeline optimized via Bayesian methods to predict $T_{\mathrm{C}}$ from chemical composition. Trained on ~35,000 compounds, the model achieved MAE $\approx$ 41 K and RMSE $\approx$ 81 K under 10-fold cross-validation, and provided case analyses of rare-earth intermetallics and magnetic phase diagrams \cite{jung2024machine}.

Graph Neural Networks (GNNs) have also emerged as powerful models for predicting materials properties, offering high accuracy at reduced computational cost compared to DFT by learning directly from atomic connectivity \cite{xie2018crystal}. Interestingly, GNNs can also be applied to chemical formulas alone, without explicit structural data. For example, Xue and Hong represented compositions as fully connected element graphs, achieving high accuracy across multiple properties (e.g., $R^2 = 0.95$ for bulk modulus) via ensemble and multi-task learning \cite{xue2024materials}. Similarly, Xie et al. encoded stoichiometries as weighted element graphs, reaching state-of-the-art accuracy (e.g., MAE = 0.0241 eV/atom for formation enthalpy) while demonstrating strong sample efficiency, uncertainty quantification, and transfer learning \cite{goodall2020predicting}.
GNNs have also outperformed XGBoost in melting temperature prediction (RMSE = 160 K) \cite{hong2022melting}, and their features further improve regression and classification tasks when combined with XGBoost \cite{deng2021xgraphboost}.

Together, these advances show that ML, and GNNs in particular, enable accurate, composition-based predictions of physical properties, reducing reliance on costly experiments and intensive first-principles simulations, and accelerating materials discovery.

In this study, we began by using an automated literature-extracted dataset (Dataset 1) containing reported $T_{\mathrm{C}}$ and corresponding chemical compositions to assess the predictive performance of various machine learning algorithms based on elemental features such as electronegativity.
The steps involved in data processing, model training, and the selection of the most suitable machine learning model were thoroughly examined. During this process, numerous inconsistencies and erroneous entries were identified in the original dataset, an issue often arising from automatic data extraction methods such as Natural Language Processing (NLP). To address this, a new and validated dataset was constructed, referred to as Dataset 2, containing verified compounds and $T_{\mathrm{C}}$ values. 
The data processing steps applied to Dataset 1 were also implemented on the curated Dataset 2, followed by model training and prediction to evaluate performance using various machine learning algorithms.

We employ two distinct feature engineering strategies to construct descriptors from elemental properties: (1) stoichiometry-weighted statistical descriptors (e.g., maximum, minimum, and mean) and (2) features derived from GNNs.
The influence of individual descriptors, such as ionization energy, was also evaluated, revealing its strong predictive power. Both approaches, GNNs, and 
stoichiometry-weighted
feature engineering methods, highlighted the significance of energy-related features.
The study emphasizes the importance of high-quality, validated data and effective feature selection in building reliable ML models. Data curation significantly improved model performance, while the implementation of GNNs effectively captures atomic interactions information in materials, significantly enhancing prediction accuracy and underscoring machine learning’s potential to accelerate the discovery of magnetic materials.
\vspace{-8pt}
\section{Methods}

\subsection{Dataset Collection and Curation}
In this study, two distinct datasets were utilized.
Dataset 1 was sourced from the automatically generated database titled "Auto-generated materials database of Curie and Neel temperatures via semisupervised relationship extraction" \cite{court2018auto}, comprising ~39,820 entries extracted through automated methods from various experimental and computational literature sources.

Dataset 2, by contrast, was curated through meticulous manual review and validation of multiple reputable data sources was conducted.
These included the AtomWork database \cite{xu2011inorganic}, Springer Materials \cite{connolly2012bibliography}, Book of Magnetism and Magnetic Materials \cite{coey2010magnetism}, Auto-generated materials database of Curie and Neel temperatures via semisupervised relationship extraction \cite{court2018auto}, and the BERT-PSIE workflow dataset \cite{gilligan2023rule}.
Multiple databases were integrated to enhance accuracy and completeness by leveraging the strengths of each source, correcting inconsistencies, and filling gaps, resulting in a more reliable and comprehensive dataset.
This manual effort focused on ensuring the accuracy and relevance of reported $T_{\mathrm{C}}$ by including only verified ferromagnetic-to-paramagnetic transitions and excluding erroneous or ambiguous entries. The resulting dataset of 2,504 high-confidence samples provides a reliable foundation for machine learning, emphasizing data quality over quantity.
These data are now freely available on GitHub~\cite{FM}.

\begin{figure}[H]
    \centering
    \hspace*{-1cm} 
    \includegraphics[width=11cm, height=7cm, trim=3cm 0cm 2cm 0cm, clip]{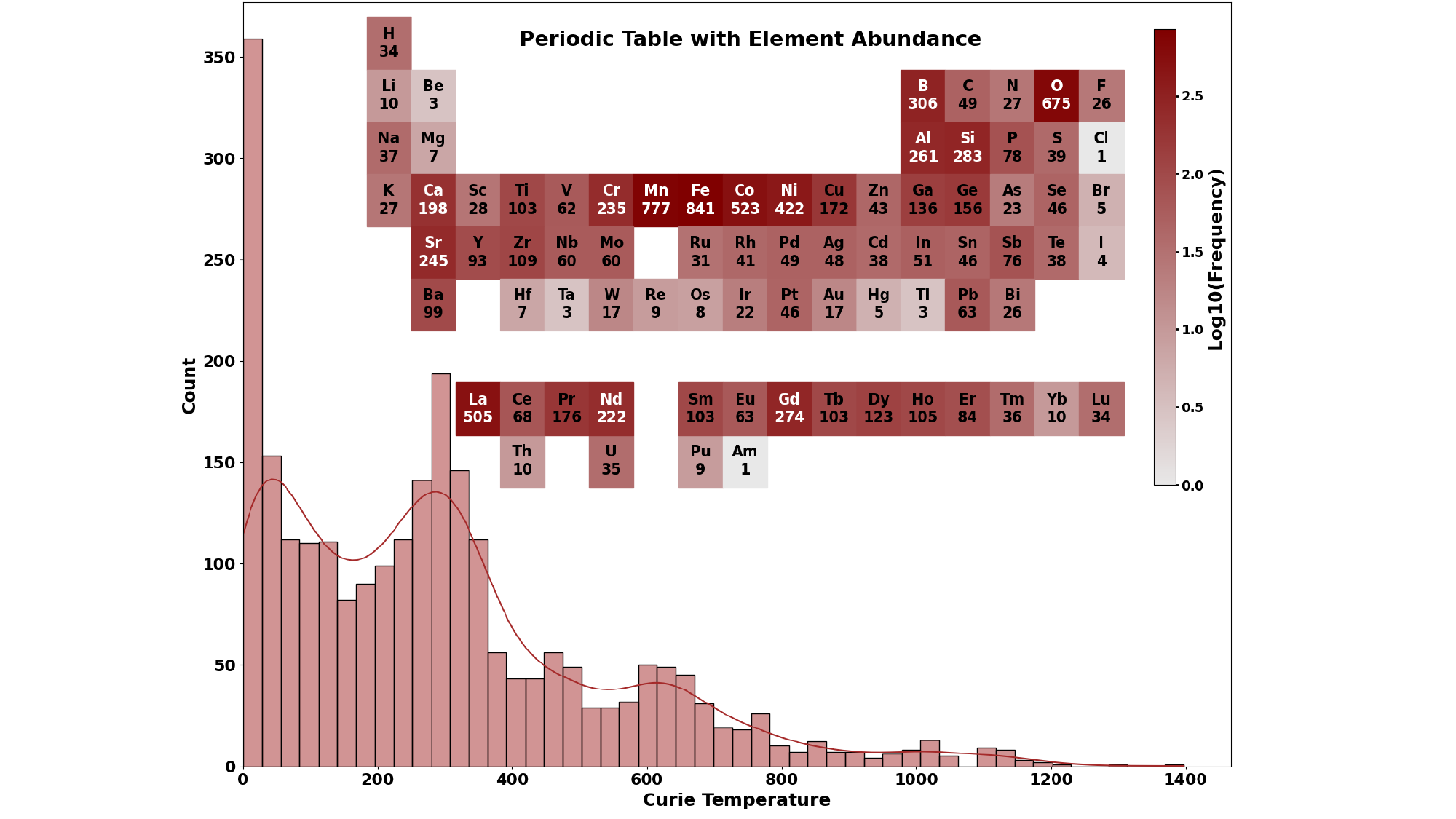}
    \vspace{-14pt}
    \captionsetup{justification=justified, singlelinecheck=false}
    \caption{Visualization of the dataset. The main panel shows a histogram of the Curie temperature distribution. The inset displays the relative elemental abundance, mapped onto the periodic table and plotted on a logarithmic scale.}
    \label{figure1}
\end{figure}

Figure~\ref{figure1} presents the essential aspects of Dataset 2. The statistical distribution of Curie temperatures across the dataset reveals that most compounds exhibit values below 400 K, while only a few exceed 1000 K, showing that stable magnetic order at high temperature is uncommon. 
Complementing this observation, the element abundance map on the periodic table highlights the central role of transition metals in magnetic materials. The most common transition metals, in descending order of abundance (from 841 occurrences of Fe to about 222 of Nd), are Fe, Mn, Co, La, Ni, Gd, Cr, and Nd. Among the non-transition elements, oxygen is the most abundant.

\subsection{Data preprocessing}

To ensure data integrity and minimize noise, extensive preprocessing was performed. First, all chemical formulas were standardized by converting fractional stoichiometries into integer representations (e.g., Ni$_{0.83}$Co$_{0.17}$ was reformatted as Ni$_{83}$Co$_{17}$). Furthermore, a simplified notation was applied where appropriate (e.g., Gd$_{95}$Ga$_{5}$ was reformulated to Gd$_{19}$Ga$_{1}$ or Gd$_{19}$Ga). 
For some compounds, multiple $T_{\mathrm{C}}$ values were reported. In such cases, the median value was used to reduce experimental variations caused by synthesis methods, doping, measurement techniques, and theoretical approximations \cite{nelson2019predicting, hwang2012emergent}.

Duplicate entries with the same compound name and $T_{\mathrm{C}}$ were first removed. To further eliminate near-duplicates, we followed the strategy of Ref. \cite{nelson2019predicting}, representing each compound by its atomic fraction vector ($v_{\mathrm{chem}}$):

\begin{equation}
v_{\mathrm{chem}}=\{x_{\mathrm{H}},x_{\mathrm{He}},x_{\mathrm{Li}},\dots\},
\end{equation}

where $x_{\alpha}$ is the atomic fraction of element $\alpha$. For example, Fe$_3$Sn$_2$ is represented as
$\{0, \dots, 0, \tfrac{3}{5}, 0, \dots, 0, \tfrac{2}{5}, 0, \dots, 0\}$,
with nonzero entries only at positions 26 (Fe) and 50 (Sn). We then compared the L$^{1}$-norms of compounds containing the same elements. If the norm difference was below 0.01, the formulas were considered redundant, and only the simpler one (fewer atoms) was retained. For instance, Ba$_{33}$La$_{67}$Mn$_{100}$O$_{300}$ and Ba$_{7}$La$_{13}$Mn$_{20}$O$_{60}$ differ by 0.008, so the latter was kept. By contrast, TiCdSe$_{2}$ and TiCd$_{3}$Se$_{4}$ differ by 0.25 and were treated as distinct compounds.

Finally, we removed entries without Curie temperature, unclear compound names, no transition metals, or reported $T_{\mathrm{N}}$ instead of $T_{\mathrm{C}}$. After cleaning, Dataset 1 was reduced to about 3,700 entries, and the manually curated Dataset 2 contained 2,500 reliable samples.

\subsection{Feature engineering}
\subsubsection{Statistical Feature Engineering using Elemental Properties (stoichiometry-weighted descriptors)} 
The performance of a machine learning model depends strongly on the choice of features. In this work, we used a broad set of features derived from elemental compositions. These descriptors were selected based on their known or expected relevance to magnetic properties. These include atomic characteristics such as periodic group and period, atomic number, valence electron count, melting point, electronegativity, polarizability, GSvolume (DFT volume per atom at T=0K), molar and atomic volume, and covalent and atomic radius, and various electronic structure parameters (e.g., band gap, Hubbard U) \cite{atomicfeatures_gitlab}. We also used a set of mathematical and statistical descriptors derived from the elemental fractions.  These include the atomic fraction vector ($ v_{\text{chem}}$) and the $L^{p}$ stoichiometry norms ($p = 1, 2, 3$), defined as ($\|x\|_{p} = \left( \sum_{i} |x_{i}|^{p} \right)^{1/p}$), which provide a compact measure of compositional spread \cite{ward2016general}. 
Additionally, we employed the stoichiometry entropy, given by
($ S = - \sum_{i} x_{i} \log(x_{i})$), which quantifies the disorder or complexity within a composition \cite{nelson2019predicting}. These descriptors, along with others, are detailed in Table~\ref{wrap-tab:1}.

\begin{table*}[htp]
\centering
\captionsetup{justification=centering,singlelinecheck=false}
 \begin{tabular}{||c c c||} 
 \hline
 Features  & Description & Dim. \\ [0.5ex]
 \hline
 \hline 
\small $L^{P}$ stoichiometry norm (p = 1,2,3) & $|| x ||_{p}  = ( \sum_{ i } || x_{i}  || ^{p} )^{1/p } $ & 3  \\ [0.5ex]
\small Stoichiometry entropy & $S  = - \sum_{ i }  x_{i}  log(x_{i})$ & 1  \\
\small Atomic fraction vector & $v_{chem}=\left\{x_{H},x_{He},x_{Li},...\right\}$ &  103 \\
\small Atomic number &  Atomic number & 6\\
\small Valence electrons & Number of valence electrons & 6\\
\small Row & Periodic table row & 6\\
\small Column & Periodic table column & 6\\
\small Molar volume & The volume occupied by one mole of each element($m^3/mol$) & 6 \\
\small Melting T & Melting temperature (K)  & 6 \\
\small Electronegativity & Electronegativity & 6  \\ 
\small (1st-10th)energy & Ionization Energy  ($E_{(1st-10th)}$(eV)) & 60\\
\small Covalent Radius & Covalent radius (pm) & 6\\
\small GSvolume & Mean DFT-computed volume of elemental solid ($A^3/atom$)  & 6\\
\small Gsbandgap & Mean DFT bandgap of elemental solid (eV)  & 6\\
\small Atomic Volume & Volume of an atom of each element ($A^3/atom$) & 6 \\
\small  U-Hubbard&  The on-site Coulomb interaction (eV) & 6\\
\small  Atomic Radius& Atomic radius (pm) & 6\\
\small  cottrell-sutton& 
A scale of electronegativity ($\chi_{CS} =\sqrt{\frac{Z_{eff}}{r}}$)
& 6\\
\small  Polarizibility& 
Static average electric dipole polarizability
& 6\\
\hline
\end{tabular}
\caption{The full list of primary features. The numerical values of the elemental properties are taken from Ref \cite{nelson2019predicting,atomicfeatures_gitlab, ward2016general,cottrell1951covalency,ralchenko2005nist}.}\label{wrap-tab:1}
\end{table*}

In addition to common features, we also employed descriptors such as the Cottrell–Sutton electronegativity 
\cite{cottrell1951covalency}, which considers both the effective nuclear charge ($Z_{eff}$) and the covalent radius to better estimate an element’s ability to attract electrons. This feature is particularly valuable for predicting and explaining trends in reactivity, bond strength, and material properties. Furthermore, we considered the first to tenth ionization energies \cite{ralchenko2005nist}, which represent how strongly electrons are bound to atoms and influence bonding, reactivity, and magnetic behavior. For elements with fewer than ten electrons, ionization energies beyond the number of available electrons were set to zero, ensuring consistency across the dataset while preserving the physical constraint imposed by electronic structure.

To construct composite features for each material, we applied six statistical operations, including minimum, maximum, mean, mode, average deviation, and standard deviation to each elemental property, weighted by the elemental stoichiometry~\cite{ward2016general,nelson2019predicting}. We refer to the resulting manually engineered features as {\it stoichiometry-weighted descriptors}.
These composite statistics capture both central tendencies and variability in the elemental attributes of each compound.
The mean and mode represent central tendencies, with the mean as the average of all values ($\bar {f} = \sum_{i=1}^{n}  x_{i}  f_{i} $) and the mode as the value of the desired feature of the element with the highest atomic fraction in the compound. This approach emphasizes the influence of the dominant element on the material’s behavior.
To quantify intra-compound variability, we calculated the average deviation, which measures the average absolute deviation from the mean ($\sum_{i=1}^{n}  x_{i} |f_{i} - \bar {f}| $), and the standard deviation, which assesses the dispersion of values around the mean ($\sqrt{{1} / {n } \sum_{i=1}^{n}  ( f_{i}- \bar {f})^{2} }$). These statistical summaries provide a comprehensive profile of the elemental distribution in each compound and are consistent with prior work in materials informatics \cite{ward2016general}.
Then all features were standardized using Min–Max scaling, transforming the raw values into a normalized range [0, 1] as follows:

\begin{equation}
X_{scaled}=\frac{X-X_{min}}{X_{max}-X_{min}}
\end{equation}

This transformation ensures that all features contribute proportionally, avoiding dominance by high-magnitude variables, and improves convergence behavior in gradient-based learning algorithms \cite{de2023choice, garcia2015data}.
Feature selection was conducted via a multi-step procedure to reduce dimensionality and eliminate irrelevant or redundant descriptors. Initially, descriptors unrelated to the dataset, such as those corresponding to elements not present in any compound, were excluded. Additionally, features with negligible correlation to the target property (e.g., "min-U-Hubbard" or "min-Gsbandgap") were discarded.

Next, a Pearson correlation analysis was performed on the remaining features. Pairs of features with a Pearson coefficient greater than 0.9 were considered highly redundant. For each such pair, the feature with the higher absolute correlation with the target variable ($T_{\mathrm{C}}$) was retained to maximize predictive relevance and interpretability.

An initial set of 257 features was refined through feature selection, resulting in 166 informative and non-redundant descriptors for Dataset 1 and 150 features per material for Dataset 2. These optimized feature sets enhance model generalization, improve computational efficiency, and preserve critical insights into ferromagnetic behavior.
\vspace{-5pt}
\subsubsection{Graph-Based Representation Learning using Elemental Properties}

In this study, we employed GNNs to learn meaningful representations of chemical compounds directly from their formulas. Specifically, we used a Graph Attention Network (GAT) architecture to capture complex interactions between elements within each compound. 
Each element within a compound is first mapped to a set of 25 standardized numerical descriptors atomic features as listed in Table~\ref{wrap-tab:1}). 
An additional value representing the element's atomic fraction in the compound is then added, creating a 26-dimensional feature vector for each element.
These enriched vectors serve as node features in a fully connected graph, where each node represents an element.
Each compound is represented as a fully connected graph (with or without self-loops), enabling the GNN to model all pairwise and higher-order interactions between elements. The GAT layers assign dynamic attention weights to neighboring nodes, identifying the most relevant elemental interactions. Through stacked GAT layers, the node features are iteratively updated, and the final node embeddings are aggregated using a weighted sum based on atomic fractions to form a fixed-length vector for the entire compound.

These learned feature vectors, which capture unary, binary, and higher-order interactions, encode the complete compositional and interactional information of the compound. The final feature vector extracted from the GNN had a dimensionality of 32, determined through hyperparameter optimization. This fixed-length vector is then used as input for various machine learning algorithms, such as XGBoost, Random Forest, and Kernel Ridge Regression to accurately predict the target property.

\subsection{Model training on dataset}
After feature extraction, the next essential phase involved selecting and training appropriate ML models for predicting $T_{\mathrm{C}}$. Model selection was guided by validation performance, with the optimal model defined as the one maximizing the coefficient of determination ($R^{2}$) metric, defined as:

\begin{equation}
R^{2} = 1 - \frac{\sum_i\left[y^{(i)}-f(x^{(i)})\right]^2}{\sum_i\left[y^{(i)}-\mu\right]^2},
\end{equation}
where $y^{(i)}$, $\mu$, and $f(x^{(i)})$ denote the true value, the mean of the true values, and the predicted value from the model using features $x^{(i)}$, respectively.

A diverse collection
of algorithms was explored, including Ridge Regression (Ridge), Kernel Ridge Regression (KRR), Support Vector Regression (SVR), Random Forests (RF), Neural Networks (NN), Extreme Gradient Boosting (XGBoost), CatBoost, and Stacking ensembles. The implementations of Ridge, KRR, SVR, and RF were carried out using Scikit-learn \cite{scikit-learn}, while neural networks were developed with Keras. Ensemble methods such as CatBoost and XGBoost, implemented using their respective libraries \cite{dorogush2018catboost,chen2016xgboost}, have demonstrated superior predictive capabilities in recent studies.
An Stacking ensemble was constructed using KRR, RF, and XGBoost as base learners, with a linear regression model serving as the meta-learner. This architecture effectively combined the strengths of individual models and enhanced generalization by reducing overfitting through diverse learning strategies.

After the feature selection step, in the modeling stage, machine learning algorithms were trained and evaluated on different feature representations. In addition to the complete feature set, two dimensionality reduction approaches, Principal Component Analysis (PCA) and correlation-based filtering (CORR), were applied to construct alternative feature spaces for model benchmarking. Furthermore, Recursive Feature Elimination (RFE) was applied only to Dataset 2 to assess its impact on model performance. RFE operates by iteratively training a model, ranking features based on their importance, and recursively eliminating the weakest ones until a predefined number of features remains \cite{zhang2022feature}. Unlike dimensionality reduction techniques such as PCA, which transform input features into new components, or correlation-based (CORR) methods that rely solely on statistical correlation, RFE evaluates feature importance within the specific context of the predictive model. This approach enables the retention of original and meaningful variables, improving both the interpretability and performance of the model. It should be noted that PCA, CORR, and RFE were not part of the primary feature selection pipeline, but were used exclusively in the modeling stage to compare algorithm performance across different feature spaces.

In the continuation of the research, we focused on applying the GNN method.
However, this approach was applied only to the reliable Dataset 2.
The features extracted from the GNN model were used as input for a diverse range of traditional machine learning algorithms. By utilizing the GNN-derived features, we aimed to further enhance the model's predictive performance and leverage the strengths of both deep learning and classical machine learning techniques.

To ensure model robustness and 
reduce overfitting, a rigorous validation protocol was adopted. The dataset was initially split into training and test sets in an 80:20 ratio. Hyperparameter tuning was conducted via K-fold cross-validation on the training set \cite{schmidt2019recent}, and final evaluations were performed on the held-out test set to estimate generalization performance reliably.
\vspace{-5pt}
\section{Results and discussions}
\subsection{Evaluating the Influence of Dataset 1 and Dataset 2 on Model Accuracy}
In order to evaluate model performance on the dataset, various machine learning algorithms were trained using the refined feature set based on stoichiometry-weighted descriptors and assessed through 5-fold cross-validation. The evaluation results on the test set of Dataset 1 are summarized in Table~\ref{wrap-tab:2}.
\begin{table}[H]
\centering
\renewcommand{\arraystretch}{1}
\begin{tabular}
{|p{1cm}p{0.4cm}p{0.4cm}p{0.4cm}p{0.4cm}p{0.4cm}p{0.4cm}p{0.4cm}p{0.4cm}p{0.5cm}|}
 \hline
 $R^{2}$  & All & C10 & C20 & C40 & C80 & P10 & P20 & P40 & P80 \\ [0.5ex]
 \hline\hline
\small Ridge & 0.39 & 0.19 & 0.27 & 0.30 & 0.35 & 0.24 & 0.31 & 0.34 & 0.38  \\ 
\small SVR & 0.55 & 0.24 & 0.31 & 0.37 & 0.45 & 0.44 & 0.50 & 0.52 & 0.55  \\
\small NN & 0.59 & 0.36 & 0.40 & 0.47 & 0.47 & 0.49 & 0.51 & 0.51 & 0.52  \\
\small RF & 0.65 & 0.52 & 0.62 & 0.61 & 0.61 & 0.57 & 0.61 & 0.60 & 0.59  \\
\small XGBoost & 0.66 & 0.49 & 0.57 & 0.57 & 0.57 & 0.52 & 0.60 & 0.60 & 0.59  \\
\small KRR & 0.64 & 0.48 & 0.57 & 0.61 & 0.62 & 0.59 & 0.57 & 0.61 & 0.61 \\
\small Stacking & 0.67 & 0.52 & 0.61 & 0.62 & 0.63 & 0.59 & 0.62 & 0.62 & 0.63  \\
\small CatBoost & 0.66 & 0.57 & 0.62 & 0.63 & 0.61 & 0.59 & 0.63 & 0.62 & 0.60  \\
\hline
\end{tabular}
\captionsetup{justification=justified, singlelinecheck=false}
\vspace{-3pt}
\caption{$R^{2}$ scores on the test set of Dataset 1 using stoichiometry-weighted descriptors with various ML algorithms through different feature reduction methods: “All” (no reduction), “C” (correlation-based), and “P” (PCA). Numbers indicate the reduced feature dimensions (e.g., P40 = PCA to 40 features).}\label{wrap-tab:2}
\end{table}
Although a robust methodology was employed, the model exhibited suboptimal performance on this database. After more investigation, these results were attributed to previously unidentified inconsistencies and erroneous entries in the dataset, including misreported $T_{\mathrm{C}}$ and the inclusion of compounds with non-ferromagnetic characteristics, such as TlCoF$_3$, Dy$_2$O$_3$, and elemental Cu. In several instances, literature sources included data corresponding to transitions other than
the ferromagnetic-to-paramagnetic phase change, such as ferroelectric-to-paraelectric, antiferromagnetic-to-paramagnetic, or ferrimagnetic-to-paramagnetic transitions.  
These discrepancies not only introduced noise but also impaired the learning process, thereby limiting predictive accuracy and generalizability.
To address these discrepancies and enhance data reliability, a systematic curation process was undertaken using multiple trusted sources. 
By cross-referencing these databases, we were able to identify and correct inconsistencies, remove non-magnetic or irrelevant entries, correct the mislabeled or ambiguous entries, and verify both the chemical composition and the reported $T_{\mathrm{C}}$ values.

Additionally, several previously unreported compounds with their corresponding Curie temperatures were discovered during this review process and were subsequently incorporated into the dataset. As a result, a high-fidelity dataset was constructed, ensuring consistency, reliability, and chemical validity.

Examples of corrected or removed entries, along with the associated sources and reasons, are summarized in 
Table~\ref{wrap-tab:Appendix} of Appendix A, illustrating the rigorous validation process undertaken to ensure the reliability and precision of the data used in this study.

The results of the evaluation of various machine learning algorithms on the test set of Dataset 2 are presented in Table~\ref{wrap-tab:3}. 
Notably, the validated dataset (Dataset 2) yielded significantly higher predictive accuracy across all models compared to the unrefined Dataset 1.

\begin{table}[H]
\centering
\renewcommand{\arraystretch}{1}
\begin{tabular}{|p{1cm}p{0.4cm}p{0.4cm}p{0.4cm}p{0.4cm}p{0.4cm}p{0.4cm}p{0.4cm}p{0.4cm}p{0.5cm}|}
 \hline

 $R^{2}$  & All & C10 & C20 & C40 & C80 & P10 & P20 & P40 & P80 \\ [0.5ex]
 \hline\hline
\small Ridge & 0.53 & 0.25 & 0.29 & 0.35 & 0.45 & 0.33 & 0.43 & 0.46 & 0.52  \\
\small SVR & 0.68 & 0.36 & 0.40 & 0.47 & 0.57 & 0.58 & 0.68 & 0.73 & 0.70  \\
\small NN & 0.80 & 0.56 & 0.70 & 0.67 & 0.71 & 0.74 & 0.83 & 0.84 & 0.83  \\
\small RF & 0.83 & 0.72 & 0.78 & 0.78 & 0.79 & 0.74 & 0.74 & 0.74 & 0.73  \\
\small XGBoost & 0.84 & 0.71 & 0.77 & 0.80 & 0.81 & 0.70 & 0.73 & 0.76 & 0.74  \\
\small KRR & 0.86 & 0.73 & 0.78 & 0.80 & 0.79 & 0.79 & 0.80 & 0.80 & 0.81 \\
\small Stacking & 0.86 & 0.74 & 0.79 & 0.81 & 0.82 & 0.78 & 0.79 & 0.80 & 0.80  \\ 
\small CatBoost & 0.87 & 0.74 & 0.81 & 0.82 & 0.81 & 0.77 & 0.77 & 0.75 & 0.77  \\
\hline
\end{tabular}
\captionsetup{justification=justified, singlelinecheck=false}

\caption{
$R^{2}$ scores on the test set of Dataset 2 using stoichiometry-weighted descriptors with various ML algorithms through different feature reduction methods: “All” (no reduction), “C” (correlation-based), and “P” (PCA). Numbers indicate the reduced feature dimensions (e.g., P40 = PCA to 40 features).}
\label{wrap-tab:3}
\end{table}

\begin{figure*}[htb]
    \centering
    \includegraphics[width=19cm, height=8.5cm]{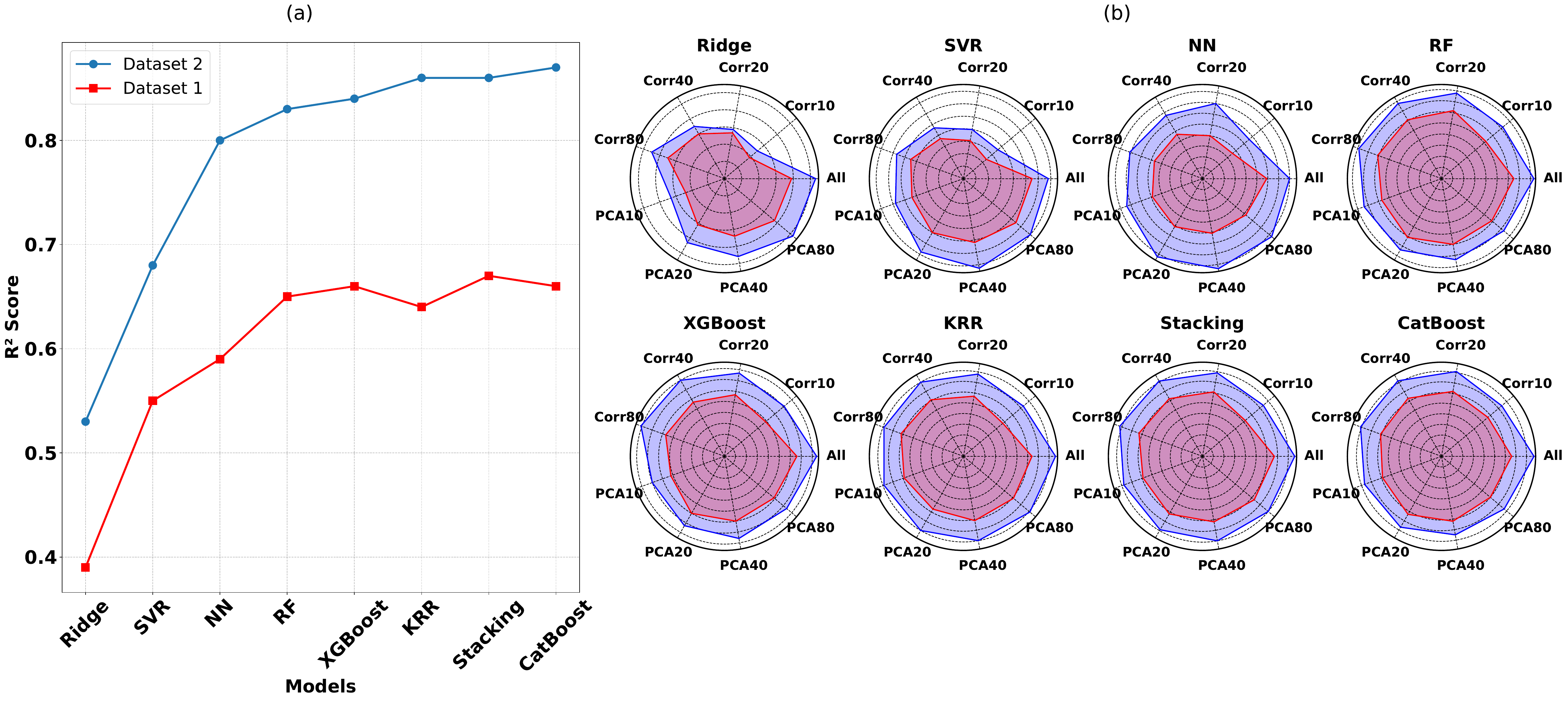}
    \captionsetup{justification=justified, singlelinecheck=false}
    \vspace{-10pt}
    \caption{(a) Line Plot of $R^{2}$ Values: Comparison of $R^{2}$ values on the test set, when all features are considered, for the eight models in the Dataset 1(red line) and Dataset 2 (blue line). 
    (b) Radar charts illustrating the effect of various feature selection methods on test set performance across different models, with the blue region representing Dataset 2 and the red region representing Dataset 1.}
    \label{figure2}
\end{figure*}

This improvement underscores the critical role of high-quality, consistent data in the training of machine learning models. Using stoichiometry-weighted descriptors, KRR and ensemble learning methods such as RF, XGBoost, Stacking, and CatBoost exhibited superior performance, with $R^2$ values reaching up to 0.87.
These models demonstrated strong predictive capability in capturing nonlinear patterns. In contrast, Ridge Regression and SVR yielded significantly lower $R^2$ scores (0.53 and 0.68, respectively), reflecting their limitations in modeling complex relationships.
Notably, although both SVR and KRR utilized kernel-based techniques, only KRR achieved high accuracy. This indicates that beyond kernel usage, the choice of regularization strategy and loss function plays a critical role in model performance.
\vspace{-1pt}
The analysis also highlighted the benefit of utilizing the complete feature set. Across all algorithms, models trained on the full set of 150 features consistently achieved higher $R^{2}$ scores, suggesting that a richer, more comprehensive representation of the data enhances predictive performance.

Figure~\ref{figure2} was generated to provide a comprehensive comparison across two datasets.
To illustrate these differences, Figure~\ref{figure2}(a) presents a line plot comparing $R^{2}$ values for models trained on all features across the two datasets. The valid dataset (Dataset 2, blue line) consistently outperforms the Dataset 1 (red line), with the CatBoost model achieving the highest accuracy. Figure~\ref{figure2}(b) further illustrates this comparison using radar charts across eight algorithms and multiple feature selection strategies (All, Corr-based, and PCA with 10, 20, 40, and 80 features). The larger enclosed area for the validated dataset (Dataset 2, blue area) relative to the Dataset 1 (red area) reflects superior overall model performance.
The distance from the center indicates the $R^{2}$ score. Higher $R^{2}$ values suggest better predictive performance.

Since Dataset 2 consistently demonstrated higher predictive accuracy, the subsequent analyses focused on Dataset 2. 

\subsection{Recursive Feature Elimination (RFE) feature selection}
In order to investigate whether model performance could be improved by selecting the most informative features, RFE was applied to Dataset 2 across four machine learning models, including Ridge, Random Forest (RF), XGBoost (XGB), and CatBoost (Cat), with results summarized in Table~\ref{wrap-tab:4}.

\begin{figure*}[!htb]
    \centering
    \includegraphics[width=19cm, height=7.5cm]{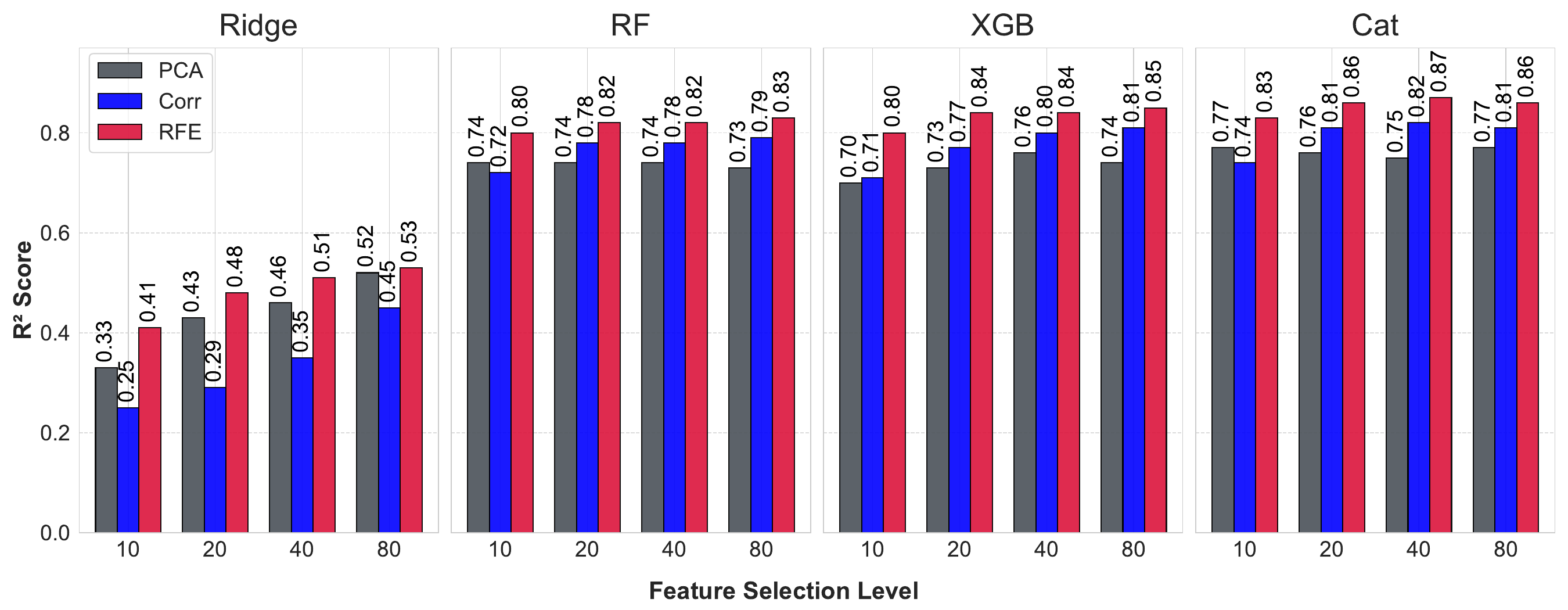}
    \vspace{-14pt}
    \captionsetup{justification=justified, singlelinecheck=false}
    \caption{Comparison of $R^{2}$ scores on the test set for different feature selection methods (PCA, Correlation-based, and RFE) across Ridge, Random Forest, XGBoost, and CatBoost models at varying feature selection levels (10, 20, 40, and 80). The bar chart visualizes performance differences.    }
    \label{figure3}
\end{figure*}

This indicates that a carefully chosen subset of features can maintain the essential predictive performance of the model while significantly reducing computational complexity. However, excessive feature elimination, as observed when selecting only 10 features, led to a marked decline in performance, as reflected by reduced $R^{2}$ values. These findings underscore the importance of a balanced approach in feature elimination, where retaining sufficient information is critical to model fidelity.
Figure~\ref{figure3} presents a comparative analysis of $R^{2}$ scores across PCA, CORR, and RFE-based feature selection methods at varying feature counts (10, 20, 40, and 80). RFE (red bar) consistently outperformed PCA and correlation-based selection, especially at higher feature selection levels and in non-linear models such as RF, XGBoost, and CatBoost.

\begin{table}[H]
\centering
\renewcommand{\arraystretch}{1.1} 
 \begin{tabular}
 {|c c c c c |} 

 \hline
 $R^{2}$  & RFE10 & RFE20 &  RFE40 & RFE80 \\ 
 \hline\hline
\small Ridge & 0.41 & 0.48 & 0.51 & 0.53    \\ 
\small RF &  0.80 & 0.82 & 0.82 & 0.83 \\
\small XGboost & 0.80 & 0.84 & 0.84 & 0.85   \\
\small Catboost & 0.83 & 0.86 & 0.87 & 0.86  \\[0.5ex]
\hline
\end{tabular}
\captionsetup{justification=justified, singlelinecheck=false}
\caption{$R^{2}$ values on the test set for different models using RFE with varying numbers of selected features (10, 20, 40, and 80), after training with 5-fold cross-validation.}\label{wrap-tab:4}
\end{table}
\subsection{Key features in Curie Temperature predictions}
Feature analysis helps to reveal how individual variables affect the predictions, while also offering valuable insight into the physical parameters most strongly associated with $T_{\mathrm{C}}$.
To quantify the contribution of key features to Curie temperature prediction, the performance of predictive models was evaluated using each feature in isolation, such as electronegativity (EN), molar volume (MV), ionization energy (Energy), and polarizability ($\alpha$), to assess the individual contributions of key features to the prediction of $T_{\mathrm{C}}$. A summary of the corresponding results is provided in Table~\ref{wrap-tab:5}.
For each feature, six statistical metrics, minimum, maximum, mean, mode, average deviation, and standard deviation, were considered to provide a comprehensive view of their predictive influence. The analysis highlighted the substantial impact of specific features on model performance, underscoring the importance of careful feature selection. 

\begin{table}[H]
\centering
 \begin{adjustbox}{width=8.60cm}
 \begin{tabular}
 {|c c c c c c |} 

 \hline
 $R^{2}$  & EN & MV & $\alpha$ &Energy & 30\_SHAP\_features \\ 
 \hline\hline
\small Ridge & 0.20 & 0.23 & 0.23 & 0.37 & 0.45  \\ 
\small SVR & 0.31 & 0.29 & 0.34 & 0.47 & 0.68 \\
\small NN & 0.58 & 0.40 & 0.53 & 0.66 & 0.79   \\
\small RF & 0.71 & 0.66 & 0.64 & 0.81 & 0.83 \\
\small KRR & 0.73 & 0.70  & 0.66 & 0.82 & 0.84  \\
\small XGboost & 0.69 & 0.66 & 0.61 & 0.84 & 0.85  \\
\small Stacking & 0.74 & 0.71 & 0.67 & 0.84 & 0.86  \\
\small Catboost & 0.75 & 0.72 & 0.68 & 0.85 & 0.86  \\[0.5ex]

\hline
\end{tabular}
\end{adjustbox}
\captionsetup{justification=justified, singlelinecheck=false}
\caption{The $R^{2}$ values of the test set illustrate the performance of various models in predicting $T_{\mathrm{C}}$ using individual features and the 30 most significant features achieved by the SHAP method.  The following abbreviations are used for key features: electronegativity (EN), molar volume (MV), ionization energy (Energy), and polarizability ($\alpha$).}\label{wrap-tab:5}
\end{table}

\begin{figure*}[!htb]
    \centering
    \includegraphics[width=14cm, height=10.5cm]{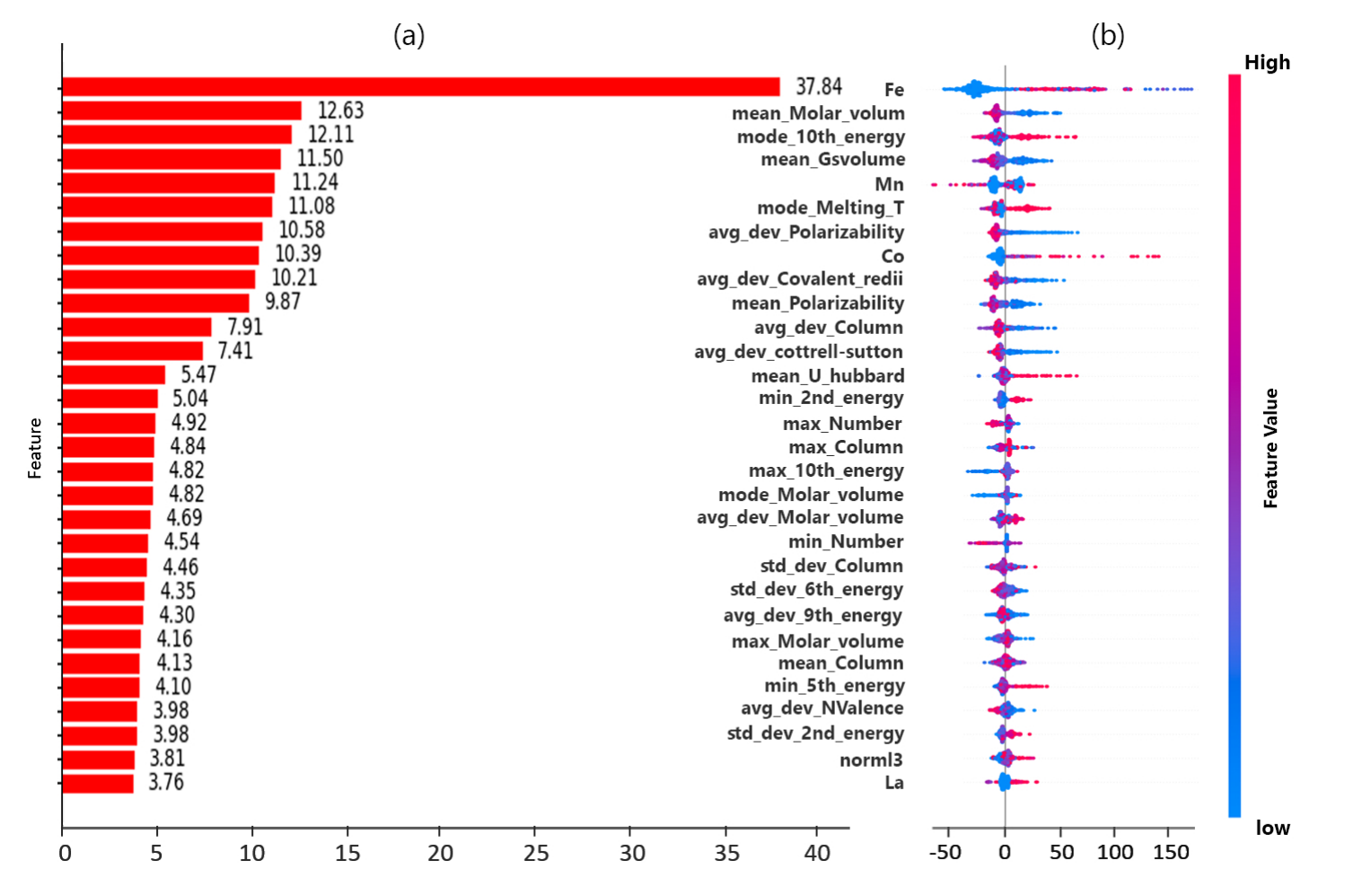}
    \captionsetup{justification=justified, singlelinecheck=false}
    \vspace{-12pt}
    \caption{Results based on the SHAP framework: 
(a) The mean absolute SHAP value of the 30 features that are identified as having the greatest contributions to the Catboost model’s prediction.
A positive SHAP value indicates that a feature has a positive contribution to the predicted output of the model.
(b) The beeswarm plot visualizes the effect of these features on the model's predictions. Each point represents an individual data instance, plotted with its corresponding SHAP value on the x-axis, while the features are listed on the y-axis, consistent with (a).
The color scheme corresponds to the original feature value, and the broadening shows the density of instances.}
    \label{figure4}
\end{figure*}

\begin{figure*}[htb]
    \centering
    \includegraphics[width=18cm, height=7.8cm]{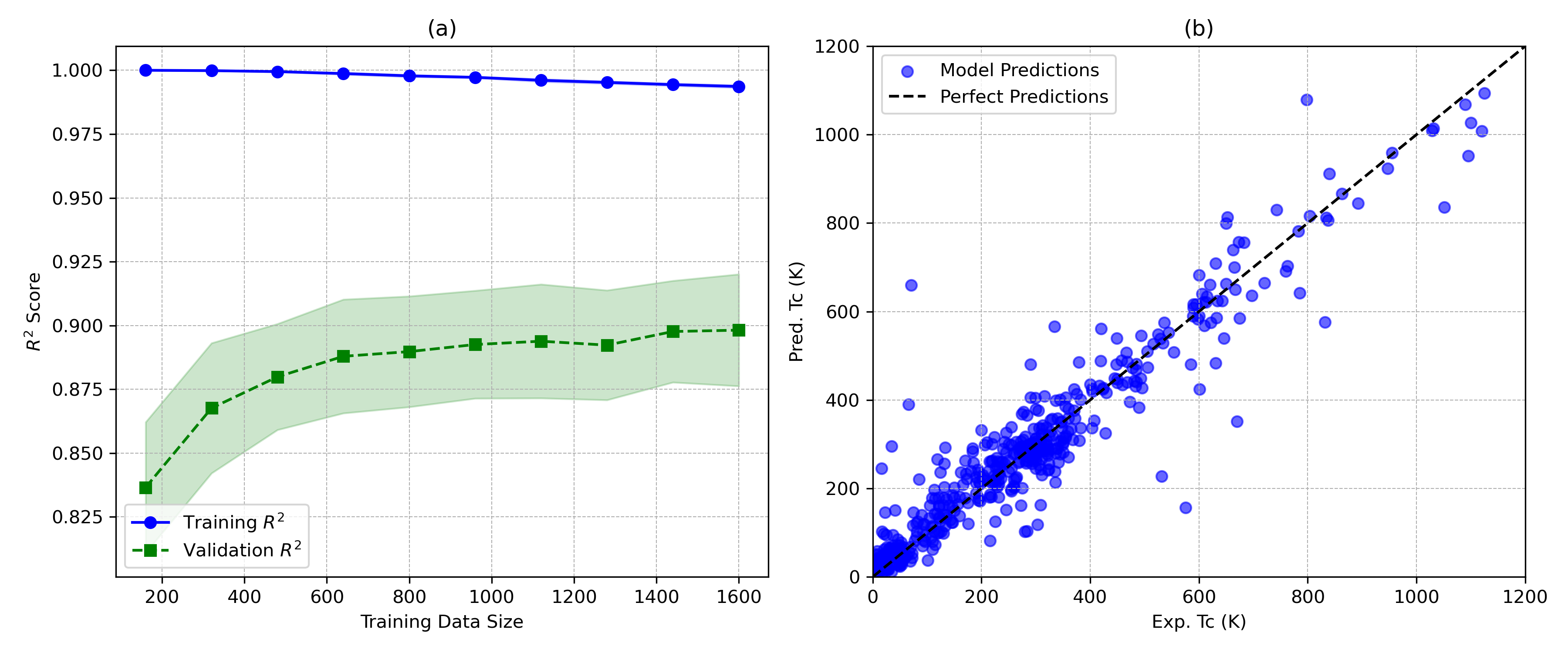}
    \caption{(a) Learning curve showing the training and validation accuracy of the Catboost model as a function of the training data size. (b) Scatter plot showing the predicted versus actual Curie temperatures on the test set using the Catboost model in the GNN method.}
    \label{figure5}
\end{figure*}

\begin{figure*}[!htb]
\hspace*{-1.5cm} 
    \centering
    \includegraphics[width=21cm, height=9.3cm]{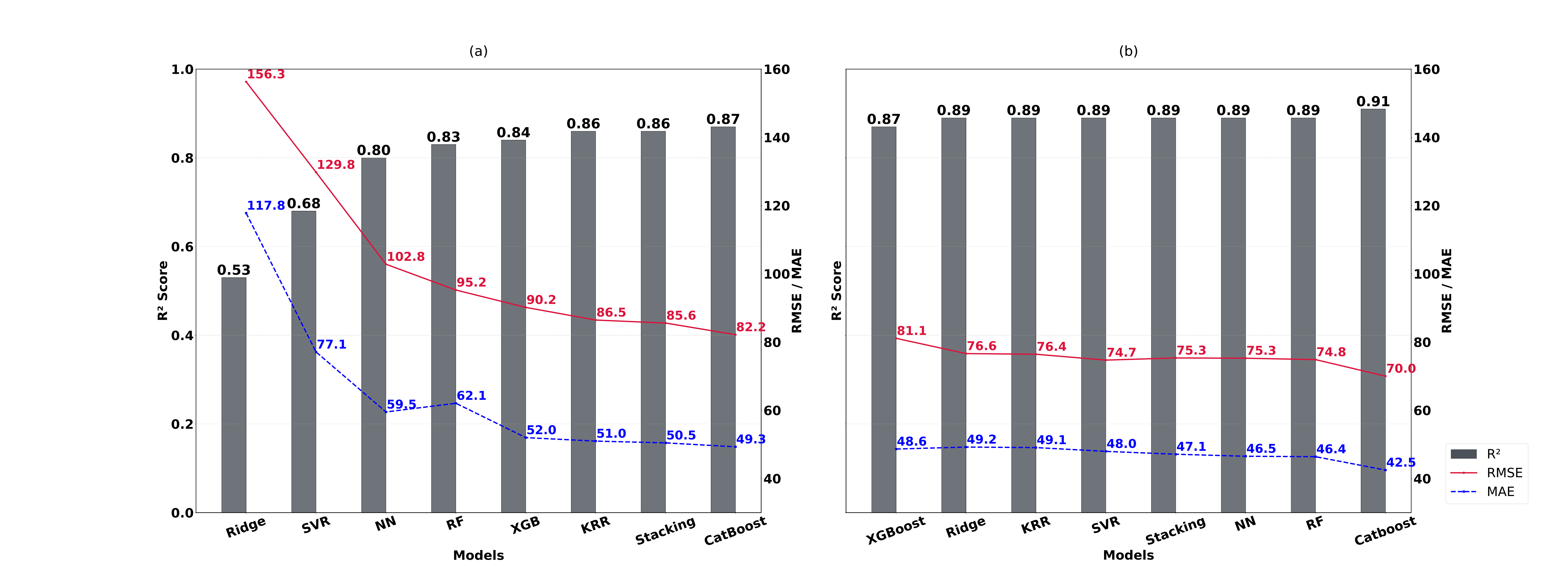}
    \captionsetup{justification=justified, singlelinecheck=false}
    \caption{Performance comparison of machine learning models using statistical features (a) and GNN-derived features (b) on the test set.  The gray bars represent the $R^{2}$ score (left y-axis), while the red line and blue dashed line indicate RMSE and MAE values, respectively (right y-axis). A higher $R^{2}$ score and lower RMSE/MAE values indicate better model performance.}
    \label{figure6}
\end{figure*}

\vspace{-7pt}
To further enhance the transparency of the model and understand the average contribution of each feature to the predictions of the model, the SHapley Additive exPlanations (SHAP) framework was employed \cite{lundberg2017unified}.
A SHAP plot is a game-theoretic approach that explain the output of any machine learning model by assigning each feature an importance value for a particular prediction. SHAP was applied to the CatBoost model, which demonstrated superior predictive performance in prior evaluations.
Figure~\ref{figure4}(a) displays the plot of the mean absolute SHAP value of the 30 features that have been identified as having the most significant contributions to the model output. Figure~\ref{figure4}(b) the beeswarm plot shows the distribution and direction of each feature’s effect on model's predictions by representing each instance as an individual data point.
Table~\ref{wrap-tab:5} also provides the results of the top 30 features identified via SHAP analysis for the test set.
In particular, the use of the 30 most influential features derived from SHAP analysis resulted in markedly improved prediction accuracy, demonstrating that targeted feature selection not only enhances model generalizability but also improves computational efficiency by reducing redundancy and mitigating the influence of irrelevant variables. Among the individual features, ionization energies emerged as particularly influential in the SHAP analysis.
The RFE-30 analysis (see Appendix B) identified 9 out of the 30 selected features as related to ionization energies, reinforcing their significance.
Therefore, it is worthwhile to evaluate machine learning models using only ionization energy-related features to assess their independent predictive power.

When ionization energies are used independently, this feature yielded strong predictive results, reinforcing its relevance to the underlying physical phenomena governing magnetic behavior.
The ionization energy was represented through ten distinct levels of the first ionization potential.
This approach provided deeper insights into the interplay between electronic structure and $T_{\mathrm{C}}$, illustrating how fundamental atomic characteristics directly influence macroscopic magnetic behavior.
These findings provide valuable guidance for the development of interpretable and high-performing machine learning models in materials science. By identifying and emphasizing features with strong predictive power, such as ionization energy, researchers can construct models that not only achieve high accuracy but also offer meaningful physical insights, thereby bridging the gap between data-driven predictions and theoretical understanding.

\subsection{GNN-Based Representation Learning Results}
In addition to the engineered feature set based on stoichiometry-weighted descriptors, a modern representation learning strategy, GNN-based representation, was also employed on Dataset 2. The GNN model was initially used as an independent predictor (R² = 0.78), but its accuracy was insufficient for direct prediction. 
However,  the learned graph embeddings from the GNN captured rich compositional information. Therefore, instead of using the GNN solely for prediction, it was used as a feature extractor to generate graph-level descriptors. These GNN-derived features were then used as inputs to classical machine learning models (e.g., CatBoost, XGBoost), significantly improving performance.

A comprehensive set of 25 elemental features together with elemental fractions (in total 26 features, see Methods) was used for node feature construction in the GNN, and the resulting representations were evaluated across machine learning models (Table~\ref{wrap-tab:6}, first column). 

According to the strong performance of ionization energies in the previous section, their role was further investigated by incorporating different subsets of ionization energies with elemental fractions as GNN inputs. The GNN-derived representations were then used in CatBoost to assess the contribution of electronic states to material behavior.

Accuracy improved to $R^2$ = 0.62, 0.75, and 0.85 with the inclusion of elemental fractions and the first one, first three, and first ten ionization energies, respectively.
\begin{table}[H]
\centering
 \begin{adjustbox}{width=9cm}

 \begin{tabular}
 {|c c c |} 

 \hline
 $R^{2}$  & \makecell{25 elemental properties \\ + element’s atomic fraction}  
          & \makecell{10 ionization energies \\ + element’s atomic fraction} \\ 
 \hline\hline
 Ridge &  0.89 & 0.87 \\ 
 SVR &  0.89 & 0.89 \\
 NN &  0.89 & 0.87  \\
 RF & 0.89 & 0.88 \\
 KRR &  0.89 & 0.88  \\
 XGboost  & 0.87 & 0.85  \\
 Stacking & 0.89 & 0.89  \\
 Catboost  & 0.91 & 0.85  \\[0.5ex]

\hline
\end{tabular}
\end{adjustbox}
\captionsetup{justification=justified, singlelinecheck=false}
\caption{Comparison of GNN-based feature representations using two node feature configurations: (1) a comprehensive set of 25 elemental properties with element’s atomic fraction, and (2) the first ten ionization energy values with element’s atomic fraction. The results highlight the effect of node-level feature design on model performance.}
\label{wrap-tab:6}
\end{table}
 
Lower-order ionization energies primarily capture valence bonding, whereas higher-order values provide information on inner-shell stability and multi-electron interactions.
Using all ten ionization energies along with elemental fractions enhanced feature representation, offering a more accurate link between electronic structure and magnetic properties, and yielding a robust descriptor for predicting Curie temperatures. The corresponding results are reported in the second column of Table~\ref{wrap-tab:6}.

Figure~\ref{figure5}(a) presents the learning curve of the CatBoost model with GNN-based feature representations using the full set of elemental properties, showing minimal divergence between training and validation performance as training size increases, thus indicating excellent generalization capability. The observed stability suggests that the model avoids overfitting and demonstrates robustness across various data partitions. Additionally, the scatter plot in Figure~\ref{figure5}(b) compares predicted and actual $T_{\mathrm{C}}$ values for CatBoost. The close alignment of data points along the diagonal (ideal prediction line) further confirms the model’s high accuracy in learning the underlying relationships between features and target values. Together, these results establish CatBoost as a highly effective and stable model for this predictive task.

In contrast to stoichiometry-weighted descriptors (Figure \ref{figure6}(a)), GNN-based representation learning (Figure \ref{figure6}(b)) consistently outperforms all evaluated ML models. Across various ML algorithms, $R^2$ values for GNN-derived descriptors range from 0.87 to 0.91, exceeding the best scores achieved with stoichiometry-weighted descriptors. Moreover, ML models using GNN-derived descriptors achieve lower RMSE and MAE values, indicating higher predictive accuracy and reduced errors.

\section{Conclusion}
In conclusion, this study presents the construction of a reliable and well-validated dataset of ferromagnetic materials, underscoring the pivotal role of high-quality data in enhancing the predictive accuracy of machine learning models. 
Beyond classical approaches based on statistical feature engineering of elemental properties, we show that GNN-based representation learning improves the accuracy of ML algorithms, achieving higher $R^2$ scores and lower RMSE and MAE values.
A comprehensive evaluation of multiple algorithms revealed that ensemble approaches, such as RF, XGBoost, Stacking, and especially CatBoost with an $R^2$ value of 0.91, obtained using GNN-based feature representations, consistently outperformed other methods in the prediction of magnetic transition temperatures.  
By analyzing key elemental features, we find that ionization energy alone can yield nearly the same results as using the full set of elemental features. Incorporating fundamental material properties like ionization energy helps develop accurate and generalizable machine learning models for materials science applications. 
Collectively, these findings emphasize the importance of integrating advanced representation learning techniques, such as GNNs, with careful feature selection.

\section*{Declarations}

\subsection*{Funding}
M.~A.\ acknowledges financial support from the Iran National Science Foundation (INSF) under Project No.~4002648.  
N.~R.\ and A.~R.~O.\ acknowledge support from the Russian Science Foundation under Grant No.~19-72-30043.

\subsection*{Acknowledgements}
Not applicable.

\subsection*{Data availability}
The Dataset 2 is available at 
\url{https://github.com/malaei/Magnetic_data/FM}

\subsection*{Author Contributions}
A.A.O.\ performed all calculations, generated all figures, gathered and curated the dataset, and drafted the initial manuscript.  
A.M.\ supervised the research, contributed to the project design, and substantially revised and edited the manuscript.  
A.R.O.\ provided critical feedback and valuable suggestions that helped shape both the project and the manuscript.  
All authors reviewed and approved the final version of the manuscript.
\vspace{-15pt}
\subsection*{Competing Interests}
The authors declare no competing interests.

\printbibliography

\end{multicols}
\newpage
 \appendix
 \label{sec:appendixA}
\section*{Appendix A}
    \renewcommand{\arraystretch}{1.1} 
\captionsetup{justification=raggedright,singlelinecheck=false}
    \begin{longtable}{|p{2.9cm}|p{1.9cm}|p{6.7cm}|p{4.9cm}|}
        \hline
        \textbf{Composition} & 
        \textbf{T($K$)}& 
        \textbf{Transition or Temperature type} & 
        \textbf{Reference (DOI address)}  \\
        \hline

MnGaFe$_2$ & 240 & Antiferromagnetic-Ferromagnetic & 10.1016/j.jallcom.2013.01.145 \\
\hline
CrAgS$_2$	& 42   &  Antiferromagnetic-Paramagnetic&  10.1016/j.jmmm.2015.12.039			\\
       \hline		

KNiF$_3$	& 250 &  Antiferromagnetic-Paramagnetic &  10.1016/S0304-8853(01)00820-4 \\				\hline
USb	& 213 &  Antiferromagnetic-Paramagnetic& 10.1016/0921-4526(95)00080-S \\				           
        \hline
CoPS$_3$	& 120 &  Antiferromagnetic-Paramagnetic &    10.1016/j.jmmm.2020.166813		\\
   \hline		
Tb$_2$Au	& 50   &  Antiferromagnetic-Paramagnetic&  10.1016/j.jmmm.2017.07.070		\\
        \hline		
TlCoF$_3$	& 20   &  Antiferromagnetic-Paramagnetic&  10.1016/j.jmmm.2015.04.028	\\
         \hline
CeAl$_3$Cu &	2.5 &  Antiferromagnetic-Paramagnetic & 	10.1016/j.jmmm.2006.12.001	\\	
\hline	
MnPS$_3$ &	78	&  Antiferromagnetic-Paramagnetic& 10.1016/S0304-8853(97)00666-5		\\
        \hline	
Gd$_3$Pd$_4$	& 18  &  Antiferromagnetic-Paramagnetic & 10.1016/0304-8853(92)90246-K	\\
         \hline	
FeSiRu$_2$	& 270   & Antiferromagnetic-Paramagnetic &  10.1016/j.jmmm.2013.07.058		\\
         \hline
Fe$_2$IrPt	 & 505 &  Antiferromagnetic-Paramagnetic & 10.1016/S0304-8853(00)01102-1 \\			\hline	
GdCu$_5$	& 12.5 &  Antiferromagnetic-Paramagnetic & 10.1016/j.jmmm.2009.05.024 \\					\hline
Mn$_3$AIC & 272 & Antiferromagnetic-Paramagnetic & 10.1103/PhysRev.125.1893  \\
        \hline
Dy$_2$O$_3$	& 8 & 
Antiferromagnetic-Paramagnetic & 10.1016/S0304-8853(02)01549-4 \\				\hline	
LaMnO$_3$	& range[65,300] & 
Antiferromagnetic-Paramagnetic & \url{https://github.com/Songyosk/CurieML}
 \\				\hline	
GdMn$_2$	& 86 &  Ferrimagnetic-Ferrimagnetic  & 10.1016/0022-5088(80)90142-3	\\              
        \hline	
GaFeO$_3$ &	210	  & Ferrimagnetic-Paramagnetic & 10.1016/j.jssc.2011.07.006	\\
               
        \hline	

TbAl$_7$Fe$_5$	& 242 & Ferrimagnetic-Paramagnetic & 10.1016/j.jallcom.2013.05.003 \\		\hline	
Fe$_2$CoO$_4$ & 86  & Ferrimagnetic-Paramagnetic & 10.1039/c3ce41663a  \\
        \hline

Ba$_{10}$AlFe$_{119}$O$_{190}$	&	725	&  
Ferrimagnetic-Paramagnetic & 10.1016/j.jmmm.2016.10.140	\\
               
        \hline
LiVCuO$_4$ & 383 & Paraelectric-Ferroelectric & 10.1007/s10948-015-3058-x \\
\hline
TiPbO$_3$ & 	423	& Paraelectric-Ferroelectric	& 10.1016/S1044-5803(00)00085-1	\\     
\hline
Tb$_4$Bi$_3$	& 192  &  Paramagnetic curie temperature  & 10.1016/j.jallcom.2007.06.035	\\
               
        \hline	
YbInCu$_4$	& 42	& Valance transition temperature & 10.1016/j.physb.2005.01.046		\\            \hline
EuGa$_4$ &	6 &	Superconductor phase transition & 10.1016/j.jmmm.2003.11.005			\\               
        \hline
Fe$_2$CuO$_4$ & 	116 & Verwey transition
& 10.1016/j.jallcom.2020.156898	\\         
        \hline	
	
Mn$_2$NiB	& 345    & Martensitic transformation & 10.1016/j.jmmm.2017.08.006	\\             
        \hline		
Mn$_{33}$In$_{17}$Ni$_{50}$ &	188  &  Martensitic transformation & 10.1016/j.jmmm.2018.10.036	\\     
        \hline
Mn$_{39}$Ni$_{50}$Sn$_{11}$ &	400  &  Martensitic transformation   & 10.1016/j.jmmm.2018.11.071\\  
        \hline

Fe$_3$O$_4$ & 43 & Unreported formula & 10.1016/0921-4534(90)90215-Z \\
        \hline								
Pt & 95.7  & Unreported formula
         & 10.1016/j.jssc.2009.01.027  \\
        \hline
AlN$_2$	& 45  & Unreported formula and temperature & 10.1016/j.jmr.2020.106683		\\        
        \hline	
Sm$_2$Co &	3   &  Unreported formula and temperature & 10.1016/S0304-8853(98)01028-2	\\
               
        \hline
MnFe$_2$Sn	& 2 &  Unreported temperature, ($T_{c(correct)}$=1012)&  10.1016/j.jmmm.2020.166426		\\  
        \hline
MnGaAs & 400  & Incorrect temperature, ($T_{c(correct)}$=110) & 10.1016/j.ssc.2020.114172  \\
        \hline

MnFeP	& 300  & Incorrect formula  &  10.1016/j.physb.2006.11.010		\\        
        \hline
Fe$_2$O$_3$ &	1098 &	Incorrect formula & 10.1016/j.jcrysgro.2010.10.181	\\
               
        \hline	
MnB	 & 20 & Incorrect formula & 10.1016/j.jallcom.2004.10.042		\\
               
        \hline	
CrCuO$_2$ &	120 & Incorrect formula 	& 10.1016/j.jallcom.2011.04.153		\\
               
        \hline	
Hf(FeGe)$_6$ &	353 &	Incorrect formula
 & 10.1016/S0925-8388(02)00669-2	\\             
        \hline				
Sr$_{40}$Li$_3$Cu$_{20}$(BrO)$_{40}$ &	46 & Incorrect formula
 & 10.1016/j.physc.2006.03.070	\\
\hline
PCOF &	3	& Abbreviation of Pr$_2$FeCrO$_6$ & 10.1016/j.jssc.2019.120903	\\
               
        \hline		
YBCO &	84	& Abbreviation of YBa$_2$CuO$_{7-\delta}$,

Superconductor phase transition & 10.1016/S0921-4526(98)00046-5		\\
               
        \hline	    

\caption{Examples of materials with uncommon Curie temperature transitions\cite{coey2010magnetism, jung2024machine, court2018auto, xu2011inorganic, connolly2012bibliography, gilligan2023rule}}

\label{wrap-tab:Appendix}
\end{longtable} 
\newpage
\section*{Appendix B: CatBoost Performance Using RFE-30 Features}

\subsection*{Selected Features (RFE-30)}
The following 30 features were selected using Recursive Feature Elimination (RFE) with the CatBoost model. Features related to ionization energy are \textbf{bolded} to highlight their importance:

\vspace{8pt}

\noindent
\begin{adjustwidth}{0pt}{0pt}
[Mn, Fe, Co, norml3, avg\_dev\_U\_hubbard,  mean\_U\_hubbard, 
\textbf{min\_1st\_energy}, \textbf{min\_2nd\_energy}, \textbf{avg\_dev\_3rd\_energy}, \textbf{min\_5th\_energy}, 
\textbf{std\_dev\_6th\_energy}, \textbf{avg\_dev\_6th\_energy},\textbf{max\_9th\_energy},\\ \textbf{avg\_dev\_9th\_energy}, 
\textbf{mode\_10th\_energy}, mean\_Column, avg\_dev\_Column, avg\_dev\_Melting\_T, mode\_Melting\_T,\\
avg\_dev\_Covalent\_redii, mean\_Gsvolume, mean\_Polarizability, avg\_dev\_Polarizability, 
max\_Molar\_volume, mean\_Molar\_volume, min\_Number, max\_Number, mean\_Number,
avg\_dev\_cottrell-sutton, std\_dev\_cottrell-sutton]
\end{adjustwidth}

\subsection*{Test Set Results}
$R^2$-score-test 0.8605062932525898

RMSE-test: 85.442156

MAE-test: 54.46477243530106 

\subsection*{Discussion}
Out of the 30 selected features, \textbf{Nine features are directly related to ionization energies}, as also identified in the SHAP importance analysis. This alignment underscores the significance of ionization descriptors in model performance. Thus, a dedicated evaluation of models trained solely on ionization energy-related features is warranted to assess their individual predictive strength.

\end{document}